\documentstyle[preprint,aps,prb]{revtex}
\draft

\begin{document}

\title{Spin-$\frac{1}{2}$ Heisenberg-Antiferromagnet on the Kagom\'e Lattice:
High Temperature Expansion and Exact Diagonalisation Studies}

\author{N.~Elstner and A.~P.~Young}
\address{Physics Department, University of California, Santa Cruz, CA 95064 }

\maketitle

\begin{abstract}

For the spin-$\frac{1}{2}$ Heisenberg antiferromagnet on the Kagom\'e lattice
we calculate the high temperature series for the specific heat and the
structure factor. A comparison of the series with exact diagonalisation
studies shows that the specific heat has further structure at lower
temperature in addition to a high temperature peak at $T\approx 2/3$. At
$T=0.25$ the structure factor agrees quite well with results for the ground
state of a finite cluster with 36 sites. At this temperature the structure
factor is less than two times its $T=\infty$ value and depends only weakly on
the wavevector $\bf q$, indicating the absence of magnetic order and a
correlation length of less than one lattice spacing. The uniform
susceptibility has a maximum at $T\approx 1/6$ and vanishes exponentially for
lower temperatures.

\end{abstract}


\newpage

\section{Introduction}

For a long a time it has been speculated that low-dimensional quantum spin
antiferromagnets may have magnetically disordered ground states. Most
attention has been focused on the spin-$\frac{1}{2}$ Heisenberg
antiferromagnet (HAFM) on the square lattice due to the close relation of
this model with the problem of high temperature superconductivity. However, it
is now well established that the HAFM on this particular lattice has an ordered
groundstate \cite{Man91}.
The first system for which a disordered groundstate was proposed is the
HAFM on the triangular lattice \cite{And73}. In recent years this model has
been the subject of intensive numerical investigations. Although most results
indicate that the system remains ordered at $T=0$, it seems that the
sublattice magnetisation and the spin stiffness are significantly smaller
than for the square lattice\cite{BLP92,LRu93,SHu92,ESY93}

So far the best candidate for a magnetically disordered system is the Kagom\'e
structure: a triangular lattice with a triangular basis. The vectors of the
underlying triangular Bravais lattice are
\begin{equation}
{\bf e}_1 = 2 (1,0) \hskip 0.5in , \hskip 0.5in {\bf e}_2 = (1, \sqrt{3})
\end{equation}
while basis vectors indicating the coordinates of the three sites in the
triangular unit cell are
\begin{equation}
{\bf b}_1 = (0,0) \hskip 0.25in , \hskip 0.25in
{\bf b}_2 = (1,0) \hskip 0.25in , \hskip 0.25in
{\bf b}_3 = \left({1\over 2},{\sqrt{3}\over 2}\right) \hskip 0.5in .
\end{equation}
The units are chosen so that the nearest neighbour distance equals unity.
The structure is shown in fig.1. Because the Kagome lattice is not a Bravais
lattice, but has three sites per unit cell, the structure factor is a
$3\times 3$ matrix.

Classical spins on the Kagom\'e structure are frustrated, as they are on the
triangular lattice, but the
coordination number is smaller (4 for the Kagom\'e structure instead of 6 for
the triangular lattice). Even more important may be another difference: while
the ground state on the triangular lattice is degenerate only with respect to
global rotations in spin space, the classical ground states on the Kagom\'e
lattice have a local degeneracy, which results in a finite ground state
entropy.
Fluctuations around magnetically ordered states have a dispersionless
zero-energy mode\cite{HKB92}.

Numerical studies using series expansions\cite{SHu92} and exact
diagonalisation techniques\cite{ZEl90,LEl93} have convincingly shown the the
ground state of the spin-$\frac{1}{2}$ HAFM on the Kagom\'e lattice has no
long range magnetic order. The exact
diagonalisation studies\cite{ZEl90,LEl93} find a very rapid decay of the
spin-spin correlations indicating a correlation length $\xi$ of only about one
lattice spacing and a finite spin gap $\Delta \approx 0.25J$.
There exist a number of proposals for a disordered ground state
\cite{Els89,MZe91,Sac92}.
Large-N expansions for the SU($N$) and Sp($N/2$) generalisations
\cite{MZe91,Sac92} of the HAFM predict a ground state with spin-Peierls order
for SU($N$) or
a spin liquid for Sp($N/2$). Up to now there exist no results from numerical
calculations that confirm or contradict any of these proposals.

Apart from being a theoretical toy model for a disordered quantum spin system
there exists at least one possible realisation of the spin-$\frac{1}{2}$ HAFM
on the Kagome structure; the second layer of $^{3}$He atoms absorbed onto
graphite at a particular coverage\cite{Els89}. It is, however, possible that
this is too simple a model and a realistic description should include other
spin-exchange interactions \cite{Rog90}. Experiments on this system found a
peak in the specific heat\cite{Gre90}, but the total change in entropy per
site between $T=\infty$ and an extrapolated  value for $T=0$ accounts for only
one half of $\ln(2)$ the expected valuefor a S=$\frac{1}{2}$ system.
This suggests that
there is a large number of low-lying states, which could contribute to
additional structure, such as a second peak, in the specific heat at very
low temperatures.

Exact diagonalisation of a 12-site cluster\cite{Els89} on the Kagome lattice
and simulations using the decoupled-cell Monte Carlo technique\cite{ZEl90}
found such a peak.
Simulations of larger systems\cite{FNi93} using the forced occillator method
found only one high temperature peak and an almost linear T-dependance of the
specific heat at temperatures below this single maximum. Based on this
observation it was concluded that the double peak structure reported for the
12-site system is due to finite size effects.

Further insight in the behaviour of the system can be gained by calculating
finite temperature properties. However, the most powerful technique, Quantum
Monte Carlo simulations, breaks down due to the sign problem for frustrated
spin systems. In this paper we, therefore, follow a different approach and
present results from high temperature expansions as well as exact
diagonalisation studies.

Section 2 contains the results for the specfic heat, for which the series has
been calculated up to 16th order in $J/k_{\rm B}T$.
We analyse the series using the method of Pad\'e approximations and compare
the results with data from exact diagonalizations studies. Based on entropy
arguments we will show that the low temperature structure of the specific heat
occuring in the finite cluster calculations\cite{Els89} cannot be a spurious
finite size effect.
In the third section we present results for the spin-spin correlation
function, which has been calculated up to 14th  order in $J/k_{\rm B}T$.
Extrapolations give quantitative results down to $T \approx 0.25 J/k_{\rm B}$.
At this temperature the largest eigenvalue of the structure factor matrix
$S_{\alpha, \beta}({\bf q})$ is nearly independent of the wavevector
$\bf q$ and is less than a factor of two larger than at $T=\infty$.
Our values for the structure factor at this temperature are close to exact
results for the groundstate of a finite cluster with 36 sites\cite{LEl93}.
Our conclusions are summarized in section 4.

\section{Specific Heat}

The Hamiltonian of the Heisenberg model is given by
\begin{equation}
H = J \sum_{\left<i,j\right>} {\bf S}_i \cdot {\bf S}_j \hskip 1in J>0  \; ,
\end{equation}
where the sum runs over all pairs of nearest neighbours on the Kagom\'e
lattice. We calculated the high temperature series using a linked cluster
expansion\cite{GSH90} up to order 16 in $1/T$ (from now on we will set the
exchange
coupling $J$ and the Boltzmann factor $k_{\rm B}$ equal to unity). The series
coefficients are given in table 1. The series was extrapolated beyond its
radius of convergence by the method of Pad\'e approximants. For a power series
$F(x)$ we form the Pad\'e approximants
\begin{equation}
[L/M] = {P_L(x) \over Q_M(x)}
\end{equation}
where $P_L(x)$ and $Q_M(x)$ are polynomials in $x$ of order $L$ and $M$
respectively. The coefficients of the two polynomials are determined by the
condition that the expansion of $[L/M]$ has to agree with the series $F(x)$ up
to order $O(x^{L+M})$. Asymptotically a $[L/M]$-Pad\'e approximant has the
behaviour
\begin{equation}
\lim_{x \rightarrow \infty} [L/M]  \propto x^{L-M} \; .
\end{equation}
In case of high temperature expansions $x$ is the inverse temperature.
Because the specific heat must vanish at $T=0$ we restrict the Pade
analysis to approximants with $M>L$.

A number of approximants were obtained this way and are plotted in fig. 2.
The curves for different approximants remain consistent with each other down
to $T=0$. Furthermore, our findings agree with results from simulation studies
on small clusters (of up to 18 sites)\cite{FNi93}. In particular there is
only one peak at $T \approx 2/3$.
We calculated the total change in entropy
\begin{equation}
\Delta S = \int_0^{\infty} {C\over T} dT
\end{equation}
and found
\begin{equation}
{\Delta S \over N} \approx 0.6 \ln (2)
\end{equation}
from integration of the Pade approximants for the specific heat. The
$[L/L]$ Pades for the entropy, which by construction go to a finite value at
zero temperature, are in agreement with this result.
Experiments on $^3$He films\cite{Gre90} absorbed on graphite reported a total
change in entropy per site of only about $\frac{1}{2}\ln(2)$ which is very
close to our results. However, this is not enough to show that
this model is appropriate to describe these experiments.
There exist other proposals\cite{Rog90} that give similar findings.

We do not, however, expect the large ground state degeneracy corresponding to
a $T=0$ entropy per site of $0.4 \ln 2$ implied by Eq. (7). Presumably
tunneling removes the high ground state degeneracy which occurs in the
classical model, leading to many low lying (but split) states for the quantum
case. This would give additional structure to the specific heat at low
temperatures where the high-T series do not converge and where the experiments
have not yet been performed.

We therefore also calculated the complete spectrum for finite systems with
N = 12, 15 and 18 sites using standard diagonalisation routines\cite{rec}.
We chose two different clusters with 18 sites. The two possibilities,
referred to as 18a and 18b respectively, are shown in fig.1. Cluster 18b is
the one used in the exact diagonalisation studies by Zeng and
Elser\cite{ZEl90} and the simulations performed by Fukamachi and
Nishimori\cite{FNi93}. Because translations are the only possible symmetry
operations of this cluster we were not able able to reduce the dimension of
the Hilbert space to a size that routines
for complete diagonalisation could be used. However, we calculated a large
number of the low-lying eigenvalues by a Lanczos algorithm\cite{CWi}, which
gives the accurate specific heat at the low temperatures of interest.
Due to the larger symmetry of cluster 18a, we were able to calculate all
eigenvalues of this particular system as well as for the N=12 and 15 clusters.

Results for the
specific heat are presented in fig. 3 together with one of the Pad\'e
approximants for comparison. The two different methods, high temperature
expansions in the thermodynamic limit and exact diagonalisation for finite
clusters, are in excellent agreement with each other down to $T\approx 0.3$,
i.e. below the high temperature peak. Thus, results from the finite
cluster calculations are already in the thermodynamic limit for $T\ge 0.3$

At lower temperatures, however, the two method give completely different
results. The finite clusters with an even number of
sites develop a sharp peak in the specific heat, while the 15 site system
has at least a clear shoulder. In simulations\cite{FNi93} the shoulder for the
15 site system appeared to be much less significant and no peak was observed
in the 18 site system. This led to the conclusions that the low temperature
peak reported earlier\cite{Els89} was a finite size effect. Here we see that
there {\it is} additional structure in the specific heat at low
temperatures for larger sizes. We expect that some structure persists in the
thermodynamic limit leading to a vanishing entropy as $T\rightarrow 0$. The
precise form of this structure, e.g. broad shoulder or second peak, however,
is difficult to deduce from our results, because the finite size corrections
show a large even-odd asymmetry. Exact diagonalisation studies of finite
clusters\cite{LEl93} find that the lowest triplet (quadruplet) for finite
systems with an even (odd) number of sites $N$ has an excitation energy that
remains finite in the thermodynamic limit resulting in a spin gap
$\Delta \approx 0.25$. The low-lying states that give rise to the additional
structure in the specific heat are singlets (doublets) for even (odd) $N$.

\section{Structure Factor and Uniform Susceptibility}

In addition to the specific heat we also calculated the high temperature series
for the spin-spin correlations. Only 14 terms were determined for this series,
because many more clusters contribute to this expansion than for the specific
heat. From the series for the correlations in real space we calculated the
structure factor.
As already mentioned in the introduction the Kagom\'e structure is not a
Bravais lattice so the structure factor is a $\rm3\times 3$ matrix,
given by
\begin{equation}
S_{\alpha,\beta}({\bf q}) = \sum_{\bf R} \exp[-i{\bf q}\cdot({\bf R}
+ {\bf b}_{\beta} - {\bf b}_{\alpha}) ] \left< S^z({\bf b}_{\alpha}) \cdot
S^z({\bf R} + {\bf b}_{\beta}) \right>
\end{equation}
where $\bf R$ is summed over Bravais lattice vectors formed from ${\bf e}_1$
and ${\bf e}_2$ in eq. (1) and the \linebreak ${\bf b}_{\alpha}$ ,
$\alpha = 1,2$ or $3$ are given by eq. (2). Harris et al.\cite{HKB92} showed
that up to sixth order in $1/T$ the largest eigenvalue of this matrix is
independent of the wavevector $\bf q$. This effect is due to the geometrical
properties of the lattice. This degeneracy is broken in 7th order for quantum
spins and in 8th order for the classical case. For certain
values of the wavevector with high symmetry the eigenvectors are independent
of $T$ so the eigenvalues can be obtained as series in $1/T$. The results for
the largest eigenvalue at ${\bf q} = 0$, ${\bf q} = {2\pi\over 3} (1,0)$
(corner of the Brillouin zone) and ${\bf q} = {\pi\over\sqrt{3}} (0,1)$
(center of an edge of the Brillouin zone) are given in table 1.
It is remarkable that the values for the different wavevectors
differ by only about 5\% indicating only weak dispersion.
The $\bf q$-value with the largest coeffcient in the structure factor series
changes at each order of the expansion, see table 1.
This questions the conclusion in the classical limit\cite{HKB92}, for which a
tendancy towards selection of ${\bf Q} = {2\pi\over 3} (1,0)$ was reported
based on an eight term series. The next order of the expansion may already
change that.

For wavevectors not at high symmetry points in the Brillouin zone we first
calculated the Pad\'e approximants for all nine matrix elements of
$S_{\alpha,\beta}({\bf q})$, evaluated them at a fixed temperature and then
diagonalized the matrix. In fig. 4 we present results of a scan through the
Brillouin zone at temperature $T=0.25$. For this data we first made a
transformation to the new variable $u=\tanh(f/2T)$ where $f=1/8$ and performed
the Pad\'e analysis in the series for this variable. We found that the
transformed series behaved better then the original one. Figure 4 shows that
the  largest eigenvalue is nearly independant of $\bf q$. Although the
different $[L/M]$ Pad\'es have a weak dispersion, there is no clear
tendancy towards selection of a particular $\bf q$-value. We estimate that the
largest eigenvalue of $S_{\alpha,\beta}({\bf q})$ is given by
\begin{equation}
4S_{max}({\bf q},T=0.25) = 1.72 \pm 0.04 \; ,
\end{equation}
and any dispersion is smaller than the error bars. For comparison: the
corresponding result at $T=\infty$ is $4S_{max} = 1$.

In order to check whether our results give a correct picture of the
low temperature magnetic properties we calculated the structure factor
for the groundstate of a finite cluster with 36 sites. The groundstate of
this cluster has wavevector ${\bf q}=0$. The correlations
in real space are given in ref. 12. For the 36 site system the
allowed wavevectors are ${\bf q} = (0,0)$ , ${\pi\over3} (1,0)$ ,
${2\pi\over 3} (1,0)$ , ${\pi\over\sqrt{3}} (0,1)$ and others related by
symmetry to these. The results
are plotted in figure 4. The agreement between the results obtained from
the series analysis at $T=0.25$ and the ground state of the 36 site cluster
is remarkable. It indicates that the results of the series analysis
remain qualitatively unchanged down to $T=0$, i.e. there is no divergence of
the structure factor which would indicate magnetic order.
The structure factor at $T=0$ is roughly a factor of two larger than at
$T=\infty$ and has only moderate dispersion, indicating a correlation length
of less than one lattice spacing.

Results for the uniform susceptibility are plotted in figure 5 using the
series for the susceptibility in table 1.
We show only results for finite cluster with an even number of sites $N$.
As already mentioned at the end of the previous section the lowest states for
a finite system with an odd number of sites are doublets which results
in a strong finite effect for the susceptibility: $\lim_{T\rightarrow 0}\chi =
\frac{1}{4T} \frac{1}{N} $. Results from
the high temperature expansion and from finite cluster calculations again
agree down to $T\approx 0.3$. Above this temperature the susceptibility is
significantly smaller than the Curie-Weiss susceptibility. At low temperatures
the susceptibility appears to vanish exponentially, because the low-lying
states are all singlets. As mentioned earlier the spin gap $\Delta$ is
estimated\cite{LEl93} to be $\Delta \approx 0.25$.
Because the the $S^z = \pm 1$ components of the lowest triplet give
identical contributions to the susceptibility one expects the maximum in $\chi$
to occur at temperature $T_{max} \approx \Delta/(1+\ln(2))$.
This can be verified for the finite clusters with $N$ sites by taking the
finite size value $\Delta(N)$.
The maximum for the susceptibility is given by $\chi_{max} = 0.14 - 0.15$ with
only a rather weak $N$ dependance.

Experiments on $^3$He films absorbed on graphite find a cusp in the
susceptibility near \linebreak 1 mK\cite{SLCS93}. Unfortunately our data do
not allow to give quantitative estimates for the position and peak value of
the susceptibility at such low temperatures, because the series expansion is
no longer reliable and the finite cluster calculation suffer from finite size
effects. It is very likely that in this temperature regime, $T\ll 1$, the
spin-$\frac{1}{2}$ HAFM is no longer an appropriate model for $^{3}$He films
and additional interactions have to be taken into account\cite{Rog90}.

\section{Conclusions}
We have presented results from high temperature expansions and exact
diagonalisation studies for the specific heat and the structure factor of the
Heisenberg antiferromagnet on the Kagom\'e lattice. Our main result is that
the specific heat has additional structure a second peak or possibly a
shoulder at very low temperatures in addition to a peak at higher temperature
($T\approx 2/3$). This had been conjectured earlier\cite{Els89} but
subsequent simulations\cite{FNi93} had seemed to contradict it.
Our conclusions that
this unusual low temperature behaviour exists comes from the follwing
observations: high temperature series expansions, which do not find a second
peak, obviously cannot account for a significant amount of entropy. At
temperatures $T<0.25$ finite cluster calculations show much more structures
(a peak for even number of sites $N$, a significant shoulder for $N$ odd)
in the specific heat. For $T\ge 1/4$ the results of both methods agree,
showing that in this regime the finite cluster calculations give the correct
results in the thermodynamic limit. Therefore, the specific heat below
$T<0.25$ has to be much larger than previously reported to get the entropy
right, i.e. zero as $T\rightarrow 0$. Thus, what is seen in
the finite cluster calculations is not a spurious finite size effect.
It would be interesting to measure the specific heat at somewhat lower
temperature to see if additional structure appears.

The series expansions for the structure factor at $T=0.25$ give results that
are very similar to the results for the ground state in exact diagonalisation
studies\cite{LEl93}. Both indicate that correlations fall of rapidly with
distance. The structure factor is less than a factor of two larger the
infinite temperature value even at $T=0.25$ . Any possible dispersion is to
weak too be clearly identified. All this indicates, that the Heisenberg
antiferromagnet on the Kagom\'e lattice has no long range magnetic order.

For the susceptibility we find a maximum around $T\approx \frac{1}{6}$ and an
exponential drop at lower temperatures.

\acknowledgements
One of us (NE) acknowledges support by the Deutsche Forschungsgemeinschaft.
The work of APY is supported by NSF grant DMR 9111576.

\newpage

\centerline{\large Table 1}
\vskip 0.5in

 \begin{tabular}{|r|r|r|} \hline
  \phantom{J}N\phantom{J} &  \phantom{W}$4C$\phantom{W} & \phantom{W} $4\chi$
\phantom{W} \\ \hline
  0\phantom{J} &   0\phantom{J} & 0\phantom{J} \\
  1\phantom{J} &   0\phantom{J} & 4\phantom{J} \\
  2\phantom{J} &  48\phantom{J} & -32\phantom{J} \\
  3\phantom{J} &   0\phantom{J} & 192\phantom{J} \\
  4\phantom{J} &  -9792\phantom{J} & -384\phantom{J}  \\
  5\phantom{J} &   0\phantom{J} & -1280\phantom{J} \\
  6\phantom{J} &   4106880\phantom{J} & -155136\phantom{J} \\
  7\phantom{J} &   -5193216\phantom{J} & 2711184\phantom{J} \\
  8\phantom{J} &   -2927834112\phantom{J} & 56705024\phantom{J} \\
  9\phantom{J} &    11470159872\phantom{J} & -1716811776\phantom{J} \\
  10\phantom{J} &   3193027983360\phantom{J} & -47711784960\phantom{J} \\
  11\phantom{J} &   -26121748561920\phantom{J} & 2004747075584\phantom{J} \\
  12\phantom{J} &   -4944246830899200\phantom{J} & 55843726884864\phantom{J} \\
  13\phantom{J} &   70892246893658112\phantom{J} & -3367208347123712\phantom{J}
\\
  14\phantom{J} &   10284867640404983808\phantom{J} &
-88720801213743104\phantom{J} \\
  15\phantom{J} &   -234226245436710912000\phantom{J} &
\phantom{J}7723917022263705600\phantom{J} \\
  16\phantom{J} &  \phantom{J}-27538523697287477329920\phantom{J} & \\
\hline
\end{tabular}

\vskip 0.5in
For each quantity $A$ we define coefficients, $a_n$, by $A = \sum_{n=0}
{a_n\over n!} \left( \beta \over 4\right)^n$. The table shows the values of the
$a_n$ for the specific heat $C$ and the uniform susceptibility $\chi$

\newpage

\centerline{\large Table 2}
\vskip 0.5in

 \begin{tabular}{|r|r|r|r|} \hline
  \phantom{J}N\phantom{J} &  $4S_{max}({\bf q}\!=\!0)$ \phantom{W} &
\phantom{W}$4S_{max}({\bf q}\!=\!{\pi\over\sqrt{3}}(0,1))$\phantom{W} &
\phantom{W}$4S_{max}({\bf q}\!=\!{2\pi\over 3}(1,0))$\phantom{W} \\ \hline
  0\phantom{J} & 1\phantom{J} & 1\phantom{J} & 1\phantom{J} \\
  1\phantom{J} & 2\phantom{J} & 2\phantom{J} & 2\phantom{J} \\
  2\phantom{J} & 4\phantom{J} & 4\phantom{J} & 4\phantom{J} \\
  3\phantom{J} & -72\phantom{J} & -72\phantom{J} & -72\phantom{J} \\
  4\phantom{J} & -448\phantom{J} & -448\phantom{J} & -448\phantom{J} \\
  5\phantom{J} & 11872\phantom{J} & 11872\phantom{J} & 11872\phantom{J} \\
  6\phantom{J} & 122368\phantom{J} & 122368\phantom{J} & 122368\phantom{J} \\
  7\phantom{J} & -4503872\phantom{J} & -4494912\phantom{J} &
-4493120\phantom{J} \\
  8\phantom{J} & -61508608\phantom{J} & -61640704\phantom{J} &
-61749760\phantom{J} \\
  9\phantom{J} & 3088187904\phantom{J} & 3072426496\phantom{J} &
3066299904\phantom{J} \\
  10\phantom{J} & 48686666752\phantom{J} & 49120629760\phantom{J} &
49365544960\phantom{J} \\
  11\phantom{J} & -3348876193792\phantom{J} & -3321064779776\phantom{J} &
-3305454742528\phantom{J} \\
  12\phantom{J} & -54711166472192\phantom{J} & -55953620766720\phantom{J} &
-56467096948736\phantom{J} \\
  13\phantom{J} & 5268689812606976\phantom{J} & 5210400086315008\phantom{J} &
5168484556144640\phantom{J} \\
  14\phantom{J} & \phantom{J}80271309635928064\phantom{J} &
\phantom{J}84156994384281600\phantom{J} &
\phantom{J}85379333782306816\phantom{J} \\
\hline
\end{tabular}

\vskip 0.5in
For each quantity $A$ we define coefficients, $a_n$, by $A = \sum_{n=0}
{a_n\over n!} \left( \beta \over 4\right)^n$. The table shows the values of the
$a_n$ for the largest eigenvalue $S_{max}({\bf q})$ of the
structure factor matrix at wave vectors ${\bf q} = (0,0)$ ,
${\bf q} = {\pi\over\sqrt{3}} (0,1)$ and ${\bf q} = {2\pi\over 3} (1,0)$

\newpage
\begin{figure}
\caption{The Kagom\'e lattice with the finite clusters used in
exact diagonalisation studies of the specific heat.}
\end{figure}

\begin{figure}
\caption{[L/M] Pad\"e approximants for the specific heat $C$ obtained from a 16
term high temperature expansion.}
\end{figure}

\begin{figure}
\caption{The specific heat $C$ calculated from exact diagonalisation of finite
clusters compared with results from the Pad\'e analysis of the high
temperature series. For cluster 18b only the low-lying eigenvalues have been
determined and so the results are only presented for $T<0.4$, the region where
they are valid.}
\end{figure}

\begin{figure}
\caption{The eigenvalues of the structure factor matrix at $T=0.25$. The
Pad\'e analysis was done in the new variable $u = \tanh(f/2T)$ where
$f=0.125$. The asterix marks the eigenvalues of the structure factor for the
ground state of a finite cluster with 36 sites.  The data were obtained using
the results for the spin-spin correlations of this cluster given in ref. 12}
\end{figure}

\begin{figure}
\caption{Uniform susceptibility $\chi$ from finite clusters and a Pad\'e
analysis of the high temperature series. The Pad\'e analysis was performed for
$T\ln(T\chi)$. The data shown were obtained from the $[7/5]$ Pad\'e
approximant. The Curie-Weiss susceptibility $4\chi_{\rm CW} = 1/(T-\Theta)$
with $\Theta = -1$ is shown for comparison. The first two terms in the high
temperature series expansions for $\chi$ and $\chi_{\rm CW}$ are equal, so
$\chi \simeq \chi_{\rm CW}$  at sufficiently high $T$.}
\end{figure}


\newpage

\begin{thebibliography}{50}

\bibitem{Man91} E. Manousakis, Rev. Mod. Phys. {\bf 63} 1 (1991) and
references therein

\bibitem{And73} P. W. Anderson, Mat. Res. Bull. {\bf 8} 153 (1973)

\bibitem{BLP92} B. Bernu, C. Lhuillier, and L. Pierre, Phys. Rev. Lett {\bf
69}, 2590 (1992)

\bibitem{LRu93} P. W. Leung and K. J. Runge, Phys. Rev. B{\bf 47}, 5861 (1993)

\bibitem{SHu92} R.R.P. Singh and D. Huse, Phys. Rev. Lett {\bf 68}, 1766 (1992)

\bibitem{ESY93} N. Elstner, R. R. P. Singh and A. P. Young, Phys. Rev. Lett.
{\bf 71}, 1629 (1993)


\bibitem{HKB92} A. B. Harris, C. Kallin and A. J. Berlinsky, Phys. Rev. B{\bf
45} 2899 (1992)

\bibitem{ZEl90} C. Zeng and V. Elser, Phys. Rev. B{\bf 42} 8436 (1990)

\bibitem{LEl93} P. W. Leung and V. Elser, Phys. Rev. B{\bf 47} 5459 (1993)

\bibitem{Els89} V. Elser, Phys. Rev. Lett. {\bf 62} 2405 (1989)

\bibitem{MZe91} J. B. Marston and C. Zeng, J. Appl. Phys. {\bf 69} 5962 (1991)

\bibitem{Sac92} S. Sachdev, Phys. Rev. B{\bf 45} 12377 (1992)

\bibitem{Rog90} M. Roger, Phys. Rev. Lett. {\bf 64} 297 (1990)

\bibitem{Gre90} D. S. Greywall and P. A. Busch, Phys. Rev. Lett. {\bf 62} 1868
(1989); D. S. Greywall, Phys. Rev. B{\bf 42} 1842 (1990)

\bibitem{FNi93} K. Fukamachi and H. Nishimori, preprint

\bibitem{GSH90} M. P. Gelfand, R. R. P. Singh and D. A. Huse, J. Stat. Phys.
{\bf 59} 1093 (1990)

\bibitem{rec} W. H. Press, B. P. Flannery, S. A. Teukolsky and W. T.
Vetterling, {\sl Numerical Recipes, The Art of Scientific Computing}, Cambridge
University Press 1986

\bibitem{CWi} J. K. Cullum and R. A. Willoughby, {\sl Lanczos Algorithms for
Large Symmetric Eigenvalue Computations}, Birkhauser, Boston 1985

\bibitem{SLCS93} M. Siqueira, C. P. Lusher, B.P. Cowan and J. Saunders, Phys.
Rev. Lett. {\bf 71} 1407 (1993)

\end{thebibliography}
\end{document}